\documentclass[12pt,a4paper]{article}

\usepackage{libertine}
\usepackage[T1]{fontenc}
\usepackage[utf8]{inputenc}
\usepackage{array}
\usepackage{comment}
\usepackage{amsmath}

\usepackage[text={6.5in, 9in}, centering]{geometry}

\usepackage[printwatermark]{xwatermark}
\usepackage{wrapfig,lipsum,booktabs}

\usepackage{tabularx}
\usepackage{epstopdf}%This line makes .eps figures into .pdf - please comment out if not required.
\usepackage{subfigure}
\usepackage{graphicx}
\usepackage{etoolbox,hyperref}

\usepackage{authblk}
\usepackage[]{threeparttable}
\usepackage{multirow}

\usepackage{tikzscale}
\usepackage{tikz}
\usetikzlibrary{petri}
\usetikzlibrary{arrows, decorations.markings}
\usetikzlibrary{decorations.pathreplacing,angles,quotes}
\usetikzlibrary{automata,positioning}

\tikzset{place/.style = {circle, draw=blue!50, fill=blue!20, thick, minimum size=0.6cm},
    transition/.style = {rectangle, draw=black!50, fill=black!10, thick, minimum width=3cm, minimum height = 1cm},
    head/.style = {rectangle, draw=black!50, thick},
    neuron/.style = {rectangle, draw=black!50, thick, fill=black!10, rounded corners=0.5mm},
    blueneuron/.style = {rectangle, draw=black!50, thick, fill=blue!10,rounded corners=0.5mm},
    greenneuron/.style = {rectangle, draw=black!50, thick, fill=green!10,rounded corners=0.5mm},
    output/.style = {rectangle, draw=black!50, thick, fill=black!10},
    pre/.style =    {<-, semithick},
    post/.style =   {->, semithick}
}

\tikzstyle{VecArrow} = [thick, decoration={markings,mark=at position
   1 with {\arrow[semithick]{open triangle 60}}},
   double distance=2pt, shorten >= 5.5pt,
   preaction = {decorate},
   postaction = {draw, line width=1.6pt, white, shorten >= 4.5pt}]
\tikzstyle{innerWhite} = [semithick, white,line width=1.4pt, shorten >= 4.5pt]

\usetikzlibrary{decorations.pathreplacing,angles,quotes}

\usepackage{array}
\newcolumntype{L}[1]{>{\raggedright\let\newline\\\arraybackslash\hspace{0pt}}m{#1}}
\newcolumntype{C}[1]{>{\centering\let\newline\\\arraybackslash}m{#1}}
\newcolumntype{R}[1]{>{\raggedleft\let\newline\\\arraybackslash\hspace{0pt}}m{#1}}

\usepackage{multicol}

\usepackage{booktabs} % To thicken table lines#

\begin{document}

\title{Dynamic Spectrum Matching with One-shot Learning}
%\tnotetext[mytitlenote]{Funded by Innovate UK, project number 132200.}

\author[1]{Jinchao Liu}
\author[1,2]{Stuart J. Gibson}
\author[3]{James Mills}
\author[4]{Margarita Osadchy}
\affil[1]{VisionMetric Ltd, Canterbury, Kent, UK. E-mail: liujinchao2000@gmail.com}
\affil[2]{Department of Physical Science, University of Kent, Canterbury, Kent, UK. E-mail: S.J.Gibson@kent.ac.uk}
\affil[3]{Sandexis Ltd, Sandwich, Kent, UK. E-mail: james.mills@sandexis.co.uk}
\affil[4]{Department of Computer Science, University of Haifa, Mount Carmel, Haifa, Israel. E-mail: rita@cs.haifa.ac.il}
\date{}                     %% if you don't need date to appear

\setcounter{Maxaffil}{0}
\renewcommand\Affilfont{\itshape\small}

\maketitle

\begin{abstract}
Convolutional neural networks (CNN) have been shown to provide a good solution for classification problems that utilize data obtained from vibrational spectroscopy. Moreover, CNNs are capable of identification from noisy spectra without the need for additional preprocessing. However, their application in practical spectroscopy is limited due to two shortcomings. The effectiveness of the classification using CNNs drops rapidly when only a small number of spectra per substance are available for training (which is a typical situation in real applications).  Additionally, to accommodate new, previously unseen substance classes, the network must be retrained which is computationally intensive. Here we address these issues by reformulating a multi-class classification problem with a large number of classes, but a small number of samples per class, to a binary classification problem with sufficient data available for representation learning. Namely, we define the learning  task as identifying pairs of inputs as belonging to the same or different classes. We achieve this using a Siamese convolutional neural network. A novel sampling strategy is proposed to address the imbalance problem in training the Siamese Network. The trained network can effectively classify samples of unseen substance classes using just a single reference sample (termed as one-shot learning in the machine learning community). Our results demonstrate better accuracy than other practical systems to date, while allowing effortless updates of the system's database with novel substance classes.

% We have built a Siamese network and demonstrated its application to the problem of identifying unknown mineral samples from their Raman spectra. Convolutional neural networks (CNN) have previously been shown to provide a good solution for classification problems that utilize data obtained from vibrational spectroscopy. Their end-to-end capability allows identification from noisy spectra without the need for additional preprocessing. Superior accuracy rates for CNNs have been reported compared with other machine learning methods and with simple matching metrics. However, their effectiveness is somewhat limited by the (typically) small number of spectra per substance available to train the network. Additionally, to accommodate new, previously unseen substance classes, the network must be retrained which is computationally intensive. Here we address these issues by employing a Siamese convolutional neural network, inspired by their use in one-shot learning where typically only one example per a class (individual) is available for network training. Our results demonstrate better accuracy than nearest neighbour or large margin nearest neighbour classifiers and similar performance to CNN but without the issues associated with incorporating new classes into the model.
\end{abstract}

\section{Introduction}

Raman spectroscopy is used for the identification and quantification of solids (particles, pellets, powers, films, fibers), liquids (gels, pastes) and gases. The technique relies upon the inelastic scattering of monochromatic light, caused by interactions with molecular vibrations. A ``molecular fingerprint'' of a substance can therefore be obtained in the form of a spectrum comprising peaks that are characteristic of its chemical composition.

%incident upon an unknoprovides information about molecular vibrations and crystal structures. The characteristic fingerprinting pattern in a Raman spectrum serves as a ``molecular fingerprint'' which can be used for sample identification and quantification. Samples may be in the form of solids (particles, pellets, powers, films, fibers), liquids (gels, pastes) and gases.

Since it provides fast, non-contact, and non-destructive analysis, Raman spectroscopy has a wide range of applications in a variety of industries and academic fields. In chemistry it is used to identify molecules and study chemical bonding. In solid-state physics it is used to characterize materials, measure temperature, and find the crystallographic orientation of a sample. It has a wide range of applications in biology and medicine. Raman spectroscopy is also used for development and quality assessment in many industries such as semiconductors, polymers, pharmaceutics, and more.
Raman spectroscopy is an efficient and non-destructive way to investigate works of art. Finally, it can be used in homeland security for identifying dangerous substances.

%A Raman spectrum is a plot of the intensity of Raman scattered radiation as a function of its frequency difference from the incident radiation. This plot is a one-dimensional signal which provides ``fingerprint'' of a substance. [SG removed - repetition from above]

Pattern recognition methods can be used for automatic identification of substances from their Raman spectrum. However, most of the pattern recognition methods require preprocessing of the data. Due to this limitation, a standard pipeline for a machine classification system based on Raman spectroscopy includes preprocessing in the following order: cosmic ray removal, smoothing and baseline correction. Additionally, the dimensionality of the data is often reduced using principal components analysis (PCA) prior to the classification step.

It was shown in \cite{JLiuAnalyst}, that the Raman Spectra classification can be achieved successfully  using convolutional neural networks (CNNs).  There are three main benefits in using CNNs for vibrational spectra classification. Firstly, a CNN can be  trained to remove baselines, extract good features, and differentiate between spectra from a large number of classes in an integrated manner within a single network architecture. Thus it removes the need for preprocessing the signal. Secondly, CNN has been shown to achieve significantly better classification results than all previous methods\cite{krizhevsky2012imagenet, WangNg12, Seq2seqNN, girshick14CVPR, he2017maskrcnn, NvidiaSelfDrivingCars2016,DeepSpeech2,JLiuAnalyst}. Thirdly, the classification is very efficient in terms of computation time.

In addition to these factors, a practical  spectrum matching system
should be able to add/remove spectra from a database dynamically in real time, without the additional overheads associated with retraining the underlying classifier. Some substances might have only few or even a single sample per class. Thus a practical system should be able to provide accurate classification even when the training data is sparse. In that context, applying CNN  is problematic.  First, adding a new class to a CNN requires a change in architecture of the network (or at least its last layers) and retraining of the network\footnote{A more efficient method is to retrain the last layers and only fine tune the rest of the weights. However, the success of this approach depends on the similarity of the new class to the existing ones.} which is computationally intensive and therefore time consuming.  Second, training CNN requires many labelled samples, while using one or a few training samples per class dramatically degrades the accuracy of CNN compared to a fully trained CNN or a simpler classifier with a small training set.

Previous methods, that provide the capability of a dynamic update of the reference set with one or few samples per substance,  have tended to use very simple pattern matching algorithms. Typically, commercial systems return a short list of candidate substances ranked according to their similarity with the query spectrum according to the relative magnitudes of their hit quality index (HQI) scores. Different metrics have been used for HQI including Euclidean distance and correlation and the cosine of the angle between two spectra. The cosine similarity metric has also previously been combined with a nearest neighbour classifier ~\cite{Ishikawa2013AMC,carey2015machine}. Variations on these metrics include assigning greater weight to particularly discriminating peaks, or eliminating peaks that only occur in the query spectrum on the assumption that they are due to impurities that are of no interest. The HQI value can be affected by artefacts due to baseline and purge problems and the presence of additional peaks caused by sample contamination and is therefore susceptible to misinterpretation. Databases shipped with commercial vibrational spectroscopy instrumentation sometimes contain records for substances that simply list the positions and intensities of the peaks contained within the spectra and these could be determined from theoretical models (conversely our application is concerned with matching to reference spectra obtained empirically). Although reducing the data to peak positions allows queries to be run quickly, peak width can be important for interpretation \cite{goss1994spectral}. Multi-scale methods make use of the structures of individual peaks \cite{zhang2015multiscale,fu2016simple,tong2016recursive}. Where a sample contains an unknown mixture of substances, probabilities for the presence of each component may be obtained by, for example, a generalized linear model \cite{vignesh2012} or reverse searching using non-negative least squares \cite{zhang2014}. A reverse search ignores peaks that occur in the query spectrum but not in the reference spectrum contained in the library/database. However, these methods do not extract salient features from the data and are therefore highly susceptible to noise. Therefore preprocessing of the raw signal is required for good matching performance.

To summarize, CNN is advantages over other methods in providing  accurate,  efficient, and fully automated classification of spectra, but it requires large training sets and retraining when a substance is added or deleted from the system. In this paper, we present a system based on CNN that solves these two problems.

The problem of limited data in the CNN training can be addressed by reformulating an n-way classification task into a binary task of classifying  pairs of inputs as either the same or different classes. For this binary problem, applied to a large number of classes, even a small number of samples per class would result in a large number of training pairs (we provide further details in Section~\ref{sec:balancing_pairs}).

A special architecture, referred to as Siamese network~\cite{Bromley:1993}, has been used to determine if pairs of inputs belong to the same or different class in a number of domains (e.g., RGB images~\cite{Chopra2005LSM,Taigman:2014,FaceNet}, NIR images~\cite{cross-spectral}, speech~\cite{NIPS2011_speech} and text~\cite{NIPS2013_text}). The Siamese network architecture can be viewed as a combination of a non-linear mapping for extracting features from the input pairs with a weighted metric for comparing the resulting feature vectors. Siamese networks learn features and metrics in an integrated manner using  gradient-based learning. In this work we use a CNN for non-linear mapping as it can learn invariance and hierarchy present in the data and has exhibited unique advantages in processing spectral data.

When a Siamese network is trained on many classes, the resulting features and the corresponding learned metric are capable of generalizing beyond the classes seen in training. Thus it can be used for learning to classify unseen classes using a single training sample, termed one-shot learning.  Previous work showed the merit of using  Siamese networks for one-shot learning in character recognition and object classification (e.g. ~\cite{Chopra2005LSM,Koch2015SiameseNN}). We propose applying one-shot learning, using a Siamese network, for spectra classification and use it to build a dynamic classification system that enables online updating of a spectra database (without retraining of the system).  Specifically, for an n-way classification problem, we use a single reference sample per class, including the new classes that were not previously represented during the training of the Siamese network, and map the reference samples and the test sample to the feature space via the CNN part of the the Siamese network. Then, the nearest neighbour rule is applied using the learned similarity metric for classifying the test sample. If more samples are available per class, the comparison can be extended to k-nearest neighbors. One can also perform ranking or any other analysis of distances between the test sample to the reference set.

Our experiments show that the proposed method can perform an accurate classification of spectra even in cases where the number of classes is large and with a single or a handful of samples per class. Moreover, it allows new classes to be added or existing classes to be removed from the model in real time with no additional effort.

\def\layersep{2.5cm}
\begin{figure}
\centering
\includegraphics[width=0.825\columnwidth]{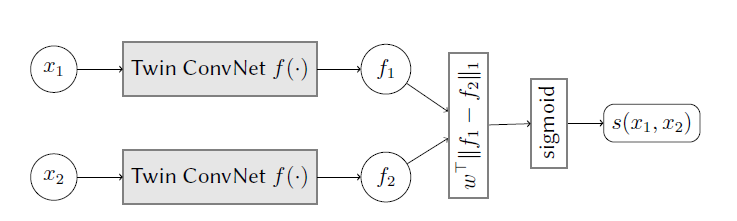}
\caption{Diagram of a Siamese network with a convolutional neural network as its twin network. $x_{1}$ and $x_{2}$ are two samples to compare, $f_{1}$ and $f_{2}$ denote their features extracted by the twin CNN. The metric in the feature space has chosen to be weighted $L_{1}$ which is learnable by adjusting $w$. Finally the network outputs a similarity measure $s(x_{1}, x_{2}) \in [0,1]$.}
\label{Fig:SiameseNetDiagram}
\end{figure}

\section{Materials and Methods}
%\section{One-shot Learning for Spectrum Recognition}
%\subsection{One-shot Learning}
\subsection{Siamese Network for One-Shot Learning}

\begin{figure*}
\centering
\includegraphics[width=0.875\textwidth]{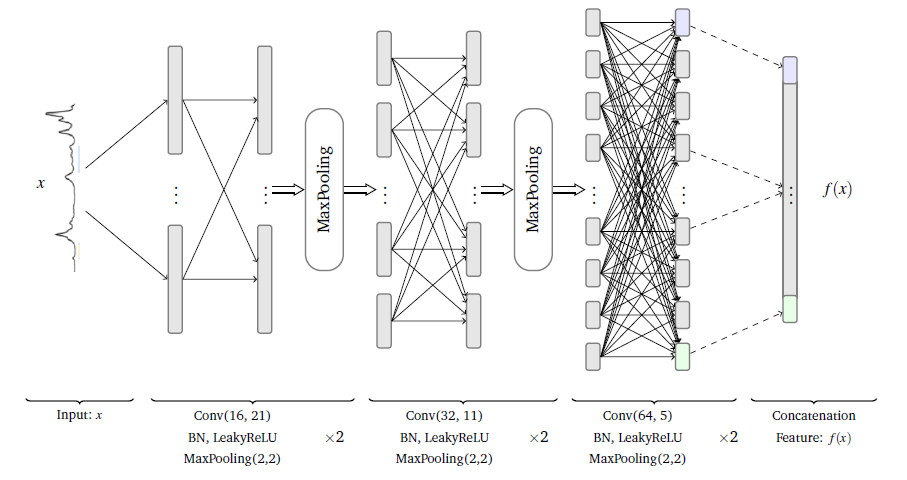}
\caption{Diagram of convolutional neural networks that are variants of LeNets\cite{Lecun98gradientbasedlearning}. This was used in our work as the twin network in Siamese networks. BN stands for batch normalization. Conv(m, n) stands for a convolutional layer with m neurons/filters of kernel size n. Maxpooling($\sigma, s$) denotes a MaxPooling layer with kernel size $\sigma$ and stride $s$. }
\label{Fig:CNNDiagram}
\end{figure*}

Siamese networks~\cite{Chopra2005LSM} consist of two (twin) networks that have exactly the same structure and identical weights. The architecture of the Siamese net used in this work is shown in Figure~\ref{Fig:SiameseNetDiagram}. We implemented the twin networks using the CNN architecture shown in Figure~\ref{Fig:CNNDiagram}. The twin network maps an input spectrum to a feature space using the same mechanism as is employed for classification using a single CNN~\cite{JLiuAnalyst} and thus enjoys all the benefits of CNN as detailed above.
The CNN architecture includes six blocks, each with a convolutional layer, followed by batch normalization, LeakyReLU non-linearity, and max-pooling. The number of feature maps is decreased in every second layer. The outputs of the last block are concatenated and flattened to form a feature vector, which is used as an input to the metric learning part of the Siamese net. Further details of the CNN architecture are shown in Figure~\ref{Fig:CNNDiagram}.

%%%%%%%%%%%%%%%%%%%%%%%%%%%%%%%%%%%%%%%%%%%%%%%%%%%%%%%%%%%%%%%%%%%%%%%%%%%%%%%%%%%%%%%%%%%%
% ConvNet diagram

\def\layersep{2.5cm}

%\footnotetext{Note that in terms of implementation, batch normalization is applied before the activation function.}

%%%%%%%%%%%%%%%%%%%%%%%%%%%%%%%%%%%%%%%%%%%%%%%%%%%%%%%%%%%%%%%%%%%%%%%%%%%%%%%%%%%%%%%%%%%%

%We use the following loss function to train the Siamese network~\cite{Koch2015SiameseNN}
%\begin{align}
%    \mathcal{L}(f,\textbf{w}) = \sum_{i,j} 1/(1 + \exp( - \textbf{w}^{\top} \| f(\textbf{x}_{i}) - f(\textbf{x}_{j}) \|_{1}))
%\end{align}
%where $f$ is a non-linear transform realized by a convolutional network (twin network) and \textbf{w} are the trainable weights that can be viewed as a metric in the feature space.
Mathematically, the model is given as follows,
\begin{align}
    s(\textbf{x}_{i}, \textbf{x}_{j}) = 1/(1 + \exp( - \textbf{w}^{\top} \| f(\textbf{x}_{i}) - f(\textbf{x}_{j}) \|_{1}))
\end{align}
where $\textbf{x}_{i}, \textbf{x}_{j}$ are a pair of samples, $f$ is a non-linear transform realized by a convolutional network (twin network) and \textbf{w} are the trainable weights that can be viewed as a metric in the feature space. We use a binary cross-entropy loss function to train the Siamese network.
\subsection{Imbalanced Positive and Negative Pairs: A Sampling Strategy}\label{sec:balancing_pairs}
One of the core problems associated with training a Siamese network is generating negative and positive pairs efficiently and sufficiently for the purpose of distinguishing both similar and dissimilar samples. Suppose there are $N$ classes, each of which has $M$ samples, the total number of positive and negative pairs are  $M(M-1)N/2$ and $M^{2}N(N-1)/2$, respectively. %It is easy to see that in general there are much more negative pairs than positive ones. Particularly,
In applications like mineral recognition, in which there are typically hundreds of different minerals, but only a handful of spectra available for each, there will be considerably more negative than positive pairs i.e. for $M \sim 10$ and $N \sim 100$, there will be roughly $5K$ positive and $500K$ negative pairs.
If we feed this ratio of training pairs into the network without proper countermeasures, positive pairs may be dominated by negative pairs during training and will therefore be under-represented in the model. As a result, the network is likely to be %twisted into only
biased towards distinguishing dissimilar pairs. An extreme case would be a classifier that only reports negative responses and fails to learn variance within the same class or similar samples.

A common way of dealing with an imbalanced dataset is to under-sample the majority class i.e. negative pairs in this case, or over-sample the minority class i.e. positive pairs, or to combine both these strategies \cite{Chawla02smote}. Here, we propose using a Bootstrapping-based strategy as follows: assume that we have $S_{m}$ positive pairs and $S_{n}$ negative pairs where $S_{m} \ll S_{n}$, for each iteration. We then sample $S_{m}$ positive pairs and an equal number of negative pairs, both with replacement. We repeat a number of iterations until the Siamese network has been trained sufficiently. It should be noted that it is crucial to sample the positive pairs with replacement, instead of generating all the positive pairs deterministically, to prevent overfitting to positive pairs.

\subsection{Training Convolutional Siamese Nets}
The Siamese network was trained using the Adam algorithm\cite{kingma2014adam} (a variant of stochastic gradient descent) for 30 iterations. The learning rate was reduced by half every 10 iterations. Xavier initialization was used to initialize the convolution layers. We applied early stopping to prevent overfitting. Training was performed on a single NVIDIA GTX-1080 GPU.

\subsection{Mineral Datasets}
We tested the proposed method on the problem of recognising  minerals using the largest publicly available mineral database, RRUFF\cite{RRUFFdataset}. We used a dataset that contains raw (uncorrected) spectra for 512 minerals. In order to investigate the performance of the tested methods on preprocessed spectra, we employed a widely-used baseline correction technique, \textit{asymmetric least squares} \cite{Paul2005}, to produce a baseline-corrected version of the dataset.
% * <s.j.gibson@kent.ac.uk> 2018-02-02T17:38:14.417Z:
%
% LMNN not defined(solved, by Jinchao)
%
% ^ <s.j.gibson@kent.ac.uk> 2018-03-02T19:02:31.592Z.

\section{Results and Discussion}
Our experiments include three parts. The first part (discussed in Section~\ref{subsect_one_shot_expr}) tests the performance of the proposed method on unseen classes. Good classification results in this experiment show the merit of the proposed method in a practical application that allows for online changes in a database (the reference set).  The second part (discussed in Section~\ref{subsect_classification_expr}) compares the performance of the Siamese network with CNN~\cite{JLiuAnalyst} in a standard multi-class classification setting. This experiment shows that the proposed method has a comparable classification accuracy with a static CNN (which requires re-training after each change in the database). Finally, we use a visualization technique (shown in Section \ref{subsect_visualization_expr}) to illustrate the ability of the network to map samples of different classes to non-overlapping clusters.
\begin{figure}
%	\centering
    \subfigure[Raw spectra where baselines can be observed.]{
	\includegraphics[width=0.475\columnwidth]{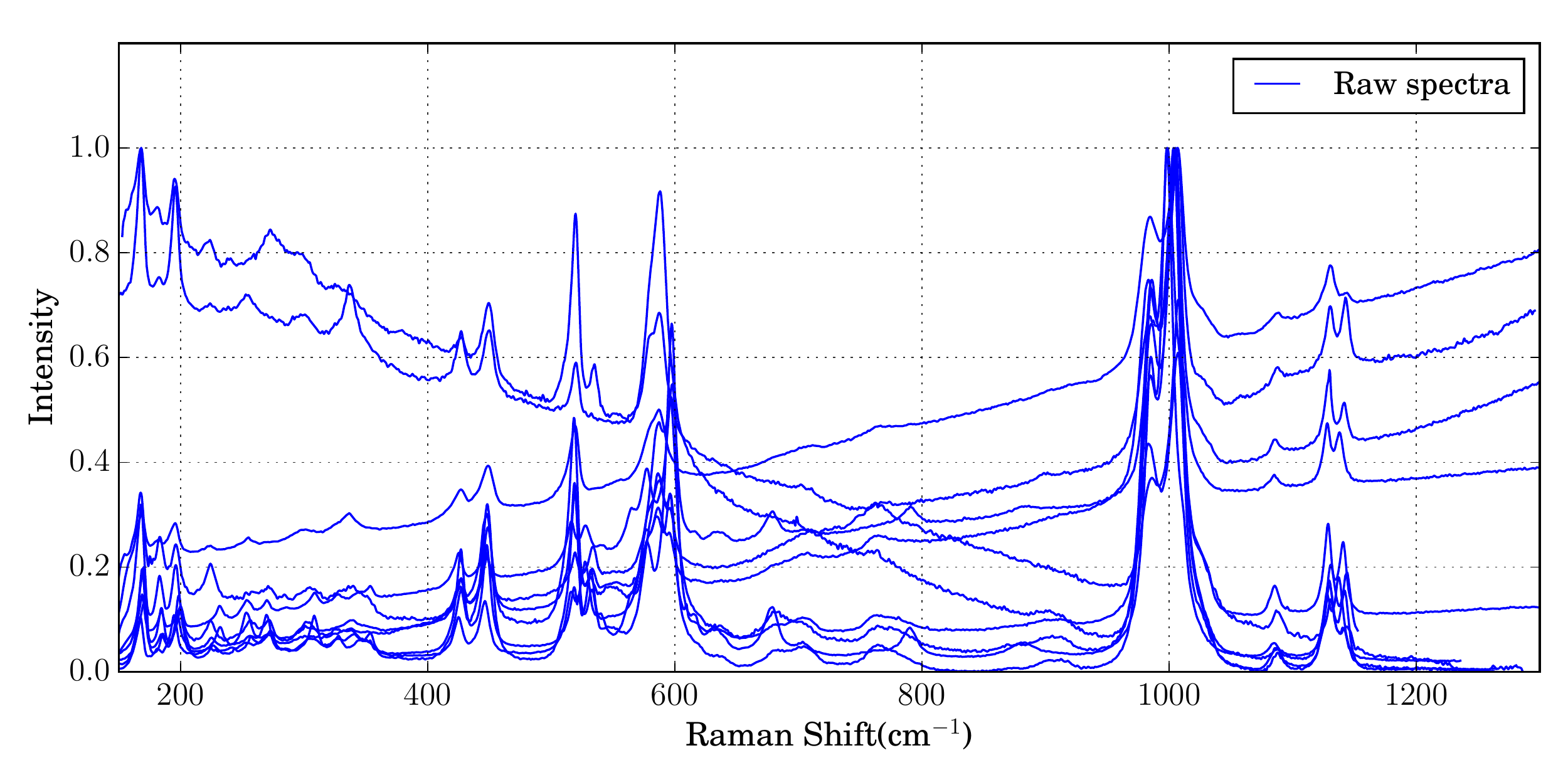}
    }
    \subfigure[Baseline corrected by asymmetric least squares]{
    \includegraphics[width=0.475\columnwidth]{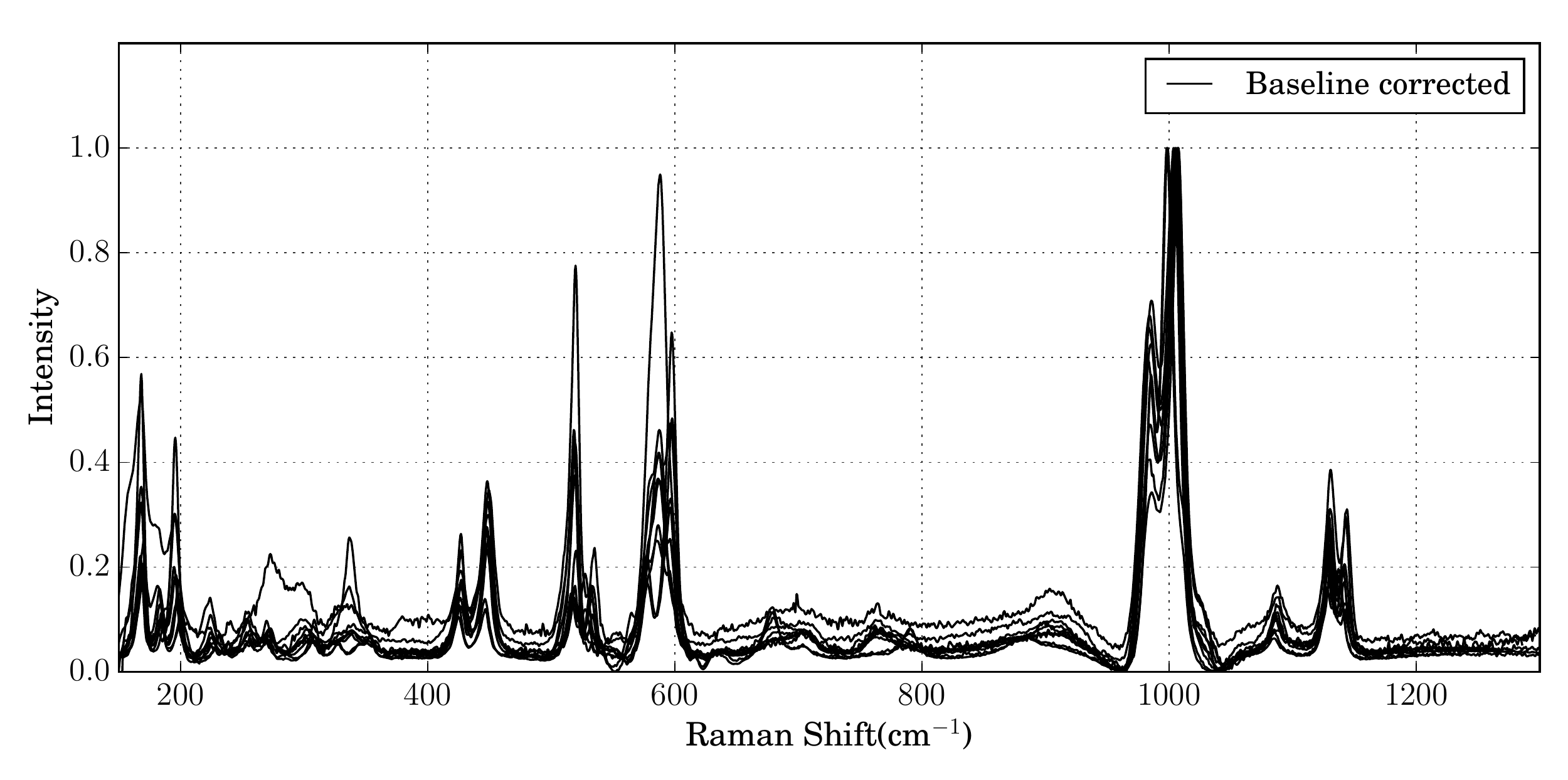}
    }
	\caption{Spectra of a mineral, \textit{hydroxylherderite}, from RRUFF raw database and corresponding baseline corrected spectra by asymmetric least squares.}
	\label{Fig:RawBaseline}
\end{figure}

%\subsection{Results and Analysis}
\subsection{One-shot Classification for Unseen Classes}\label{subsect_one_shot_expr}

\subsubsection{Evaluation Protocol}
%Similar to the test protocol of one-shot learning, we investigated the performance of the proposed method on unseen classes. But in order to simulate effectively the scenario of how this system would be used in practice, our exact protocol is defined as follows: we split all the classes (minerals) into non-overlapping training, validation and test sets, each of which contains all the samples of a set of classes.

We first investigated the proposed method on unseen classes using the following protocol: We split all 512 classes (minerals) into non-overlapping sets for training, validation, and test. We used 50\% of the classes for training, 10\% for validation, and 40\% for testing. We trained the Siamese network using all samples in the training set as discussed in Section~\ref{sec:balancing_pairs}. We validated the results for early stopping on pairs produced from the samples in the validation set. During test time, we picked at random a single sample from each class in the test set to form a reference set. We tested the classification of all other samples from the test set by searching for their best match in the reference set. Using randomization of data partition, we repeated both training and test several times to obtain statistically reliable results.

%When testing, for each class, we randomly select one sample from each test class and form a reference set. All the other samples are compared against this set to compute similarity scores.
%It should be noted that the randomness comes from both splitting the classes and forming the reference  and query sets. Therefore, the whole process is repeated a number of times to obtain statistically reliable results.
%For one shot classification, we have used 60\% of the total 512 classes for training and 40\% for testing.

\subsubsection{Results and Analysis}
We compared our method with with three others: nearest neighbour (NN) with $L_{2}$ distance, nearest neighbour with cosine similarity and large margin nearest neighbour (LMNN). Nearest neighbour with the Euclidean  distance and cosine similarity have been widely used in commercial software for spectrum matching and were therefore included for comparison. LMNN is a popular metric learning method that learns a Mahanalobis distance for $k$-nearest neighbour classification ($k$NN). A linear transform of the input space is learned such that the $k$-nearest neighbors of a sample in the training set share the same class label with the sample, while samples from different classes are separated by a large margin. In short, LMNN learns a linear transform, which is particularly beneficial for $k$NN classification. %Among the compared methods, one would expect that LMNN might outperform NN with fixed distance metrics as it is still able to

\newcolumntype{b}{X}
\newcolumntype{s}{>{\centering\hsize=.75\hsize}X}
\renewcommand{\arraystretch}{1.2}
\begin{table*}
\centering
\caption{One-shot classification accuracy ($f_{1}$-score) of convolutional Siamese nets and other compared methods on RRUFF mineral dataset with and without baseline correction. }
\label{Tab:SummaryRRUFF}
\begin{tabularx}{0.99\textwidth}{bsssc}
\toprule
Signal Type &NN($L_{2}$)  &NN(cosine) & LMNN\cite{Weinberger2009} &Siamese Net  \\ \midrule
Raw Spectra& 0.461$\pm$0.046 & 0.525$\pm$0.036 & 0.725$\pm$0.041 & 0.901$\pm$0.014  \\
Preprocessed & 0.802$\pm$0.033 & 0.833$\pm$0.030 & 0.818$\pm$0.029&  0.886$\pm$0.032 \\
\bottomrule
\end{tabularx}
\end{table*}

The results are summarized in Table \ref{Tab:SummaryRRUFF}. On raw data, NN with either $L_{2}$ or cosine similarity performed poorly, with a low accuracy rate of $\sim$ 0.5. LMNN achieved better results,  with an accuracy of 0.725.  The Siamese network significantly outperformed all tested methods and gave the highest rate of 0.901.
% * <s.j.gibson@kent.ac.uk> 2018-03-02T18:36:57.422Z:
%
% Jinchao, is it okay to call these accuracy rates? So 0.5 means no better than chance right?
%
% ^.(By Jinchao) They are not exactly accuracy rates, they are f1-scores. Here we are dealing with multi-classes classification, so the accuracy rate of random guessing would be 1/205.
%which is better than the comparable methods by a large margin.
On the baseline corrected (i.e. preprocessed) data, all previous methods performed much better (as expected).  The Siamese network achieved 0.886, which is lower than on the raw data but still significantly better than the other three methods.
% * <s.j.gibson@kent.ac.uk> 2018-03-02T18:40:46.410Z:
%
% Note: need to comment on why Siamese classification drops for preprocessed data in the discussion.
%
% ^.
%it is not a surprising that all the other three methods performed much better with the help of a baseline correction method. On the other hand, the siamese network achieved 0.886 which is still significantly better than the other three methods.

These results show that the Siamese network succeeds to learn better features and a better, problem specific, similarity metric.
%
%These results indicate that while learning a non-linear metric, the siamese network was also able to learn better feature extraction and baseline correction.
This is consistent with our previous work \cite{JLiuAnalyst}  in which end-to-end learning with a CNN resulted in classification rates for unprocessed spectra that were superior to those achieved when these spectra were analysed using pipeline methods.
%. and confirms again that convolutional neural networks can be very effective in processing Raman/vibrational spectral data.

It is worth noting that on the raw data, LMNN achieved better results than NN with both $L_{2}$ and cosine similarity (as shown in Table \ref{Tab:SummaryRRUFF}). LMNN is capable of learning a linear transform to facilitate the subsequent nearest neighbour classification. This means it has limited preprocessing capability\footnote{If one closely observes samples in Figure \ref{Fig:RawBaseline}(a), rotating the spectra clockwise or counterclockwise respectively, which is a kind of linear transform, would certainly correct the baselines to a certain degree, though not thoroughly.}, which is form of learning when compared with CNN, but certainly better than no preprocessing of the raw spectra.
% * <s.j.gibson@kent.ac.uk> 2018-03-02T18:55:27.180Z:
%
% Note to Stuart:  reword last sentence above.
%
% ^.
%It is consistent with the results where LMNN performed better than nearest neighbor without any baseline correction, but worse than deep Siamese net which learns a non-linear transform to correct the baselines.
On the other hand, on baseline corrected data, LMNN does not benefit much from the limited preprocessing and its performance was comparable with NN with the $L_{2}$ distance metric and worse than NN with cosine similarity.

\subsection{Multi-class Classification for Trained/Seen Classes} \label{subsect_classification_expr}

%We have demonstrated that convolutional Siamese nets were able to achieve excellent performance on classifying unseen classes(minerals), compared to other state-of-the-art methods for dynamic spectrum recognition. Furthermore, we also designed and ran a set of experiments to compare convolutional siamese nets with CNNs\cite{JLiuAnalyst} in terms of accuracy on trained classes.
Next we compared our convolutional Siamese network with a CNN~\cite{JLiuAnalyst} in a standard multi-class classification, in which unseen samples are classified into learned classes. We applied the Siamese network as described in Section~\ref{subsect_one_shot_expr}, but with the reference set composed of the known classes (used to train the Siamese network).
For this set of experiments, we followed the test protocol, \cite{JLiuAnalyst}, which randomly selects one sample from each class to form a test set. The remaining samples were used for training and validation. The process was repeated a number of times.

The results are summarized in Table \ref{Tab:MulticlassAcc}.
We can see that when data augmentation is used in training, the classification accuracy of the Siamese network is comparable to CNN on the preprocessed spectra and only slightly lower than CNN on the raw data. This indicates that Siamese networks can be used for a large-scale classification of spectra with almost no accuracy loss.

%When data augmentation is used in training, convolutional Siamese nets achieved $92.1\%$ on the raw data which is about $1\%$ lower than the corresponding CNN, $91.9\%$ on the preprocessed data which is comparable to the CNN. So compared to CNNs, convolutional siamese nets gained the ability of dynamically add or remove classes at the cost of a slight accuracy drop.

We investigated the effect of data augmentation on convolutioinal Siamese networks and on CNNs.
%We have also investigated how data augmentation would affect convolutioinal siamese nets and CNNs respectively.
Table \ref{Tab:MulticlassAcc} shows that the convolutional Siamese network trained with no data augmentation is able to achieve a good accuracy of roughly $85\%$ on both raw and preprocessed spectra. A trained CNN network with no augmentation, was heavily overfitted. Halving its size by keeping only one copy of each block (instead of two) resulted in significant loss in classification compared with the original network trained on augmented data, specifically, $70.5\%$ on raw spectra and $77.9\%$ on preprocessed spectra. Furthermore, a smaller CNN attained higher accuracy on the preproceessed data than on the raw spectra, which is contrary to the results of the original CNN trained on the augmented data. This suggests that the reduced CNN is not large/deep enough to learn a baseline correction well.
% * <s.j.gibson@kent.ac.uk> 2018-03-02T19:01:09.827Z:
%
% The wording of the paragraph above is problematic. I'll read it again with a fresh pair of eyes.
%
% ^.

%\newcolumntype{b}{X}
%\newcolumntype{s}{>{\hsize=.75\hsize}X}
\newcolumntype{b}{X}
\newcolumntype{s}{>{\centering\hsize=.75\hsize}X}
\renewcommand{\arraystretch}{1.2}
\begin{table*}
\centering
\caption{Classification accuracy on trained classes of convolutional Siamese nets and CNNs on the RRUFF mineral dataset, with or without baseline correction.}
\label{Tab:MulticlassAcc}
\begin{tabularx}{0.99\textwidth}{bcccc}
\toprule
Methods  & Data Augmentation  & Raw Spectra & Preprocessed Spectra  \\ \hline
\multirow{2}{*}{Convolutional Nets\cite{JLiuAnalyst}} & No$^{\clubsuit}$ & 0.705$\pm$0.031  & 0.779$\pm$0.015   \\
&Yes&0.933$\pm$0.007 & 0.920$\pm$0.008    \\ \hline
\multirow{2}{*}{Convolutional Siamese Nets}   & No    &  0.850$\pm$0.029 & 0.842$\pm$0.035    \\
  & Yes & 0.921$\pm$0.010 &  0.919$\pm$0.007&   \\
\bottomrule
\end{tabularx}
\end{table*}

Since synthetic augmentation of the data is not always easy and in some cases is not even possible, one could consider using Siamese networks for classification instead of CNN.  Siamese networks learn a binary task from pairs of samples. The number of pairs that could be obtained from a data set comprising many classes with a small number of samples in each, is large enough to prevent strong overfitting.
% * <s.j.gibson@kent.ac.uk> 2018-03-02T19:04:48.004Z:
%
% Consider putting the paragraph above in the conclusion.
%
% ^.

%While for CNNs, we had to reduce the size of the network by half and obtained $70.5\%$ on raw spectra and $77.9\%$ on preprocessed spectra. This showed that if data augmentation was not possible, siamese networks might be a better framework to train a convolutional neural network for feature  extraction than training it directly as multi-class classification problems. Careful readers may have noticed that as shown in the first row of the table CNN got higher accuracy on preproceessed than raw spectra which is contrary to the other results. This in fact indicated that the CNN was not large/deep enough to learn to preprocess the spectra correctly.

\subsection{Visualization}\label{subsect_visualization_expr}
\begin{figure}[!htp]
	\centering
	\includegraphics[width=\columnwidth]{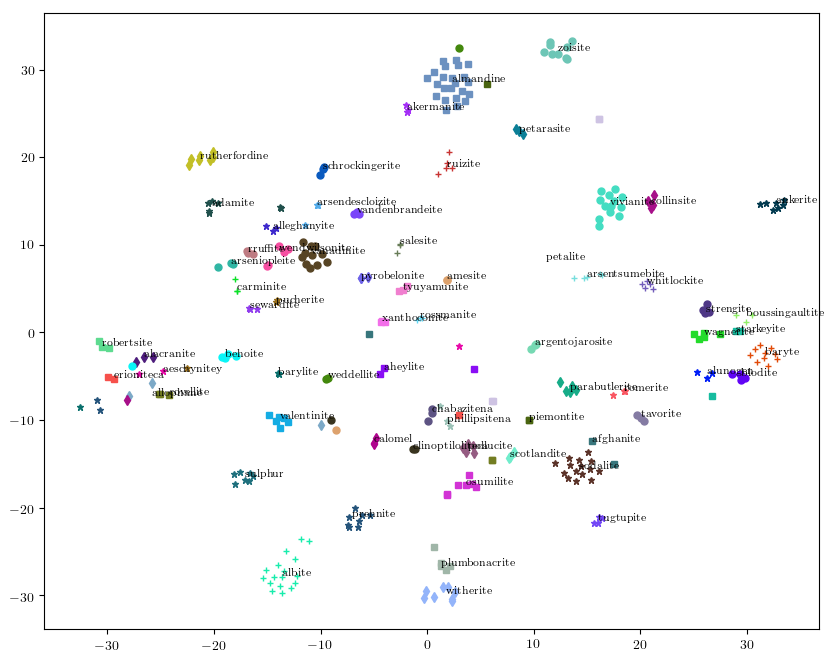}
\caption{t-SNE projection of the transformed features of some test samples. Classes are highlighted in different colors.}
	\label{Fig:tsne}
\end{figure}

We used t-distributed stochastic neighbour embedding (t-SNE)\cite{maaten2008visualizing} to visualize the feature space learned by the Siamese network. To avoid clutter in the visualization, we reduced the number of unseen classes by removing the minerals with fewer than three samples. The results depicted in Figure \ref{Fig:tsne} show that the projected samples cluster by mineral type.

%We visualized the feature space learned by the Siamese network by projecting all the transformed test samples using t-SNE\cite{maaten2008visualizing}, as shown in Figure \ref{Fig:tsne}. It can be seen that samples belong to the same mineral were projected closer to one another and samples from different minerals were projected relatively away from each other. This may serve as a graphical illustration of the non-linear mapping learned by the siamese net.

% * <s.j.gibson@kent.ac.uk> 2018-03-02T19:06:50.398Z:
%
% Jinchao, I met with Rita this week. She thought that there should be more classes present in the t-SNE image (even when single sample clusters are removed). Please can you confirm that the number of clusters shown is correct. Thanks.
%
% ^. (By Jinchao) I will double check my code for generating that figure.

\section{Conclusion and Future Work}\label{Conclusion}
In this paper, we have proposed an one-shot learning solution based on convolutional siamese networks to realize a dynamic spectrum matching system which is capable of classifying both seen and unseen classes accurately. Especially, for unseen classes/substances, the proposed system requires as few as one example per class to achieve accurate classification which allows adding classes/substances into the model dynamically. We validated the feasibility and effectiveness of our method on the largest public available mineral dataset. Although we demonstrated our method on mineral species classification using Raman spectra, the proposed framework can generalize to other kinds of spectroscopy and other applications in the computational chemistry, drug discovery, and health care etc, especially in the cases where available data for training is limited and/or frequent updates of the database are required.
%\section{Acknowledgements}
%This research was supported by Innovate UK, project number 132200.
\bibliography{RamanSpecMatching}
\bibliographystyle{ieeetr}

\end{document}